\documentclass[12pt] {article}
\usepackage{graphics}
\usepackage{epsfig}
\usepackage{psfrag}

\catcode`@=12 \topmargin -0.5in \evensidemargin 0.0in
\newcommand{\beq}{\begin{equation}}
\newcommand{\eeq}{\end{equation}}

\newcommand{\bea}{\begin{eqnarray}}
\newcommand{\eea}{\end{eqnarray}}

\begin{document} 
\thispagestyle{empty} 
\begin{flushright}
November, 2008\ 
\end{flushright}
\vspace{0.5in}
\begin{center}
{\LARGE \bf  Neutron-Anti-Neutron Oscillation: Theory and 
Phenomenology\\}  
\vspace{1.3in} {\bf
 R.~N. Mohapatra\\}
\vspace{0.2in} {\sl Maryland Center for Fundamental Physics and
Department of Physics, University of Maryland, College Park, Maryland, 
20742} \vspace{1in}
\end{center} 
 
 \begin{abstract} The discovery of neutrino masses has provided strong 
hints in favor of the possibility that B-L symmetry is an intimate 
feature of physics beyond the standard model. I discuss how important 
information about this symmetry as well as other scenarios for TeV scale 
new physics can be obtained from the baryon number violating process, 
$n-\bar{n}$ oscillation. This article presents an overview of different 
aspects of neutron-anti-neutron oscillation and is divided into the 
following parts : (i) the phenomenon; (ii) the physics,
(iii) plausible models and (iv) applications to cosmology. In particular, 
it is argued how the discovery 
of $n-\bar{n}$ oscillation can significantly affect our thinking about 
simple grand unified 
theory paradigms for physics beyond the standard model, elucidate the 
nature of forces behind neutrino mass and provide a new 
 microphysical view of the origin of matter in the universe. 
\end{abstract}

\newpage 
\baselineskip 24pt 
 
\section{Introduction}
Particle oscillations are familiar phenomena in both classical and 
quantum mechanics and have provided a wealth of information about the 
nature of matter and forces acting on them. 
 The most elementary example of a pendulum oscillation is 
caused by the force of gravity and therefore provides information about 
the strength of gravity. In the domain of quantum mechanics, two 
states close by in energy can transmute into one another 
if the Hamiltonian for the system includes a force that connects them
and if certain coherence conditions are satisfied. Turning this 
question around, one can conclude that if experimentally an oscillation 
between states, forbidden in the standard model is observed, it will then 
imply the existence of hitherto unknown 
classes of interactions. Many such instances in the 
domain of particle 
physics are known where study of particle oscillations have provided 
important clarification about the fundamental nature of 
underlying physics. Kaon and neutrino oscillations are but 
two glowing examples of this. The first in combination with the recently 
observed $B-\bar{B}$ oscillations confirmed the nature of CP violation in 
standard model. The latter has for the first time revealed the existence 
of new physics as well as new symmetries beyond the standard model. 

 The question then arises as to whether 
there are other such possibilities that need to be experimentally explored to 
probe physics beyond the standard model (SM). In this review, I argue 
that  neutron-anti-neutron ($n-\bar{n}$) oscillation provides one such 
unique example and is 
extremely timely in the aftermath of one of the fundamental discoveries 
in our field i.e. the neutrinos are massive and can provide important 
clues to 
our understanding of new symmetries and forces behind the neutrino mass 
generation.

 The $n-\bar{n}$ oscillation\cite{review} is a unique kind of oscillation 
phenomenon 
compared to kaon oscillations in that it breaks one of the 
once sacred conservations laws of physics i.e. conservation of baryon 
number which keeps all matter stable (just like kaon oscillation breaks 
strangeness quantum number). It also effectively makes neutron a Majorana 
fermion (albeit with a very tiny Majorana component), a point which is of 
some historical importance since in the original paper of Majorana, he 
contemplated neutron being the particle which is its own anti-particle.
 
The fact that $n-\bar{n}$ oscillation breaks baryon number requires 
us to first ascertain that the vast wealth of information 
available about the stability of matter from experiments and general 
cosmological observations allow for this process to have a strength that 
will make it observable with current technology. We show below that this 
is indeed the case.

Secondly as far as the physics implications go, another class of 
processes that probe baryon number nonconservation is proton 
decay\cite{nath,dusel}. They 
have been the focus of experiment as well as theory for two reasons: 
first of all, simple grand unified theories\cite{raby} based on SU(5) and 
SO(10) 
predict its
existence at a level not far from experimental capabilities; secondly, it 
used to be thought in the early 80's that these theories may also explain
the origin of matter using proton decay as one of its 
key ingredients. 

This situation however changed drastically with time due 
to two developements. 

(a) The discovery of neutrino masses was confirmed in 1998 and a 
popular paradigm for the origin of their smallness became the seesaw 
mechanism\cite{seesaw}, where new symmetry, B-L respected by then 
standard model forces had to be broken\footnote{Without any new 
symmetry, the seesaw scale would naturally be expected to be the Planck 
scale, which yields too small neutrino masses. A simple way to 
understand why the seesaw scale is lower than the Planck scale is to 
assume that it is associated with a symmetry e.g. B-L .}. Note that the 
dominant proton 
decay modes of interest $p\to e^+\pi^o$ respect the $B-L$ symmetry and is 
therefore nor suited for the study of forces responsible for neutrino 
mases. On the
other hand, $n-\bar{n}$ oscillations break B-L symmetry by two unjits 
exactly like the 
seesaw mechanism for neutrino masses does and therefore is uniquely 
suited 
for the study of the detailed nature as well as the scale of seesaw 
mechanism. There is a large class of interesting gauge models 
where indeed neutrino Majorana masses via the seesaw mechanism leads 
directly to $n-\bar{n}$ oscillations.

(b) The second development is that a new mechanism for understanding 
the origin of matter was proposed in mid 80's that did not use proton 
decay but rather 
used the seesaw mechanism for neutrino masses\cite{fuku} to generate 
matter-anti-matter asymmetry where again B-L symmetry played a key role.
Although grand unified theories such as SO(10) could incorporate the B-L 
symmetry, these two developments took a lot of the wind from the 
argument in favor of proton decay being necessarily the most 
theoretically relevant mode for baryon nonconservation. 

Furthermore, when the seesaw mechanism is 
embedded into GUT theories that include B-L symmetry, they generally tend 
to fix the scale of B-L symmetry 
breaking to be at the GUT scale, while there is no compelling 
reason for the seesaw scale to be that high. It could even 
be at the TeV 
scale without making the theory too unnatural. The seesaw scale could not 
therefore be probed by searching for proton decay whereas as
already noted, in a large class of interesting theories, $n-\bar{n}$ 
oscillation does. 

The question then arises as to whether all physics beyond the standard 
model is encapsulated in grand unified theories with neutrino masses 
being also a signal of the same GUT scale or whether the neutrino masses 
indicate an alternate path that contains rich physics far below the GUT 
scale. As we show below, $n-\bar{n}$ oscillation provides an effective 
probe of the second possibility which envisions a rich new layer of 
physics at intermediate scales that involves low B-L breaking. 
Theoretical 
models indicate that a wide range of scales, anywhere from a few 
TeV to $10^{11}$ GeV as well as the associated new physics could be 
probed by pushing the $n-\bar{n}$ oscillation 
oscillation time by three orders of magnitude over the current 
experimental lower limit. In particular, its discovery will 
 rule out simple grand unified theories or in the very least have 
significant affect on their detailed nature. Furthermore it will have  
profound effect on the nature of
physics beyond the standard model as well as provide a new microphysics 
view for the origin of matter, different from leptogenesis. It also 
provides rich new physics that can be probed at the Large Hadron 
Collider.

The bottom line is therefore that it is urgent to pursue a dedicated 
search for for $n-\bar{n}$ oscillation that can push the sub-GUT 
physics frontier to a new level. 

This review is organized into the following parts: in the first part, I 
discuss 
the phenomenon, its connection to nuclear instability ; in the next 
section, we discuss the broad character of microphysics that would lead 
to $n-\bar{n}$ oscillation and the kind of scales that can be probed by 
this. In the next section, the nature of physics beyond standard model 
probed by $n-\bar{n}$ oscillation will be discussed including a 
discussion as to why B-L is an important symmetry to be probed. In this 
section, we discuss specific and plausible gauge models for $n-\bar{n}$ 
oscillation both with and without supersymmetry. In the final section, we 
discuss the implication of observable $n-\bar{n}$ oscillation  for 
baryogenesis and show how one can have a new class of models for late 
baryogenesis after electroweak phase transition and sphaleron freeze-out.

\section{Phenomenology of $n-\bar{n}$ oscillation }
In order to discuss the main features of $n-\bar{n}$ oscillation in 
various environments, it is sufficient to consider a simple $2\times 2$ 
Hamiltonian for the evolution of an initial beam of slow moving neutrons 
 and evaluate the probability of finding an anti-neutron in this beam 
after a time $t$.
\begin{equation}
i\hbar \frac{\partial }{\partial t}\left(\begin{array}{c} n \\
\bar{n}\end{array}\right) = \left( {{\begin{array}{cc}
 {E_n } & \delta m  \\
 \delta m & {E_{\bar {n}}} \\
\end{array} }} \right)\mbox{ }\left(\begin{array}{c} n \\
\bar{n}\end{array}\right)
\label{eq:tdsch}
\end{equation}
where $\delta m$ denotes the $n-\bar{n}$ mixing which parameterizes at 
the nucleon level the underlying physics that breaks baryon number by two 
units. We will give examples of models where this can happen at the 
appropriate level in a subsequent section. Note that we have left the 
energies of the $n$ and $\bar{n}$ arbitray since they could be affected 
in different ways in the presence of an external environment such as the 
nuclear field or a magnetic field etc.

The exact expression for the probability to find an antineutron at time 
given that at $t=0$, $P_n(0)=1$ and $P_{\bar{n}}(0)=0$ is:
\begin{eqnarray}
P_{\bar{n}}(t)~=~\frac{4\delta m^2}{\Delta E^2+4\delta m^2} 
sin^2(\sqrt{\Delta E^2+4\delta m^2})t;
\end{eqnarray} 
where $\Delta E~=~E_n-E_{\bar{n}}$.
The fact that neutron decays does not have any effect on this formula as 
long as the neutron flight time is much smaller than the life time of 
neutron.

Let us consider two extreme cases of this formula which also happen to 
be interesting for experimental purposes. For $\delta m \ll \Delta E$, 
this expression reduces to: \begin{eqnarray}
 P_{ \bar{n}}\sim \left(\frac{\delta m}{\Delta E_n}\right)^2 
sin ^2 \Delta E_n t 
\end{eqnarray}
There are two special cases of this realizable in nature:
\begin{itemize} 

\item  {Case (i): $\Delta E t \ll 1$}: In this case we have,
 $P_{n\rightarrow \bar{n}}\sim (\delta m \cdot t)^2\equiv
 \left(\frac{t}{\tau_{n-\bar{n}}}\right)^2$
This case corresponds to free neutron oscillation in vacuum.

 \item { Case (ii): $\Delta E \cdot t \gg 1$}:
$P_{n\rightarrow \bar{n}}\sim\frac{1}{2} \left(\frac{\delta m}{2\Delta
E_n}\right)^2 $
This for instance corresponds to bound neutrons inside nucleus 
``oscillating'' to anti-neutrons .

 \end{itemize}

 \subsection{ Stability of Nuclei and limit on $\tau_{n-\bar{n}}$:}
As with all baryon number violating interactions, the existence of 
${n-\bar{n}}$ oscillation will lead to nuclear instability. This will 
happen 
via a conversion of neutron to an anti-neutron inside the nucleus which 
will then annihilate with the surrounding nucleons and lead to 
instability. Clearly, if the nuclear transition probability inside the 
nucleus and in vacuum were equal, this would cause all matter to 
disappear in a much shorter time than the age of the Universe and would 
be unacceptable. However, it is an experimental fact that inside the 
nucleus the neutron and the anti-neutron experience different nuclear 
potentials. In fact the difference is so large ($E_N-E_{\bar{N}}\sim 100$
 MeV or so) that the transition probability formula 
relevant for this case is that in case (ii) above and leads indeed to 
nuclear instability lifetimes of order $10^{32}$ years or so. Careful and 
detailed analysis of this question has been done by Dover, Gal and 
Richards\cite{gal1}, Alberico et al\cite{al} and more recently by Gal and 
collaborators\cite{gal2}. Below we present a crude representation of the 
above works to get the basic physics across:
\begin{eqnarray} 
P^{nuc.}_{n-\bar{n}}~\sim ~P_{n\rightarrow \bar{n}}\sim 
\left(\frac{\delta 
m}{2\Delta E_n}\right)^2.
\end{eqnarray}
where $P_{n\rightarrow \bar{n}}$ represents the free neutron oscillation 
probability in vacuum and $\delta m$ is the $\Delta B=2$ off diagonal 
matrix 
element in the $n-\bar{n}$ mass matrix in Eq. (1) and is related to the 
oscillation time $\tau_{n-\bar{n}}=\frac{h}{2\pi \delta m}$. Present ILL 
limit on $\tau_{n-\bar{n}}\geq 10^8$ sec. translates to $\delta m \leq 
6\times 10^{-33}$ GeV. A crude estimate of the nuclear instability life 
time in the presence  $n-\bar{n}$ oscillation from these intuitive 
considerations is 
$\tau_{Nucl.}\sim \left(\frac{\delta m}{2\Delta
E_n}\right)^{-2} 10^{-23}$ sec.$\simeq 10^{34}$ years. More careful 
recent 
calculations that take into account the nuclear effects\cite{gal2} are 
stated in terms of the following relation:
\begin{eqnarray}
\tau_{Nuc.}~=~R(\tau_{n\bar{n}})^2
\end{eqnarray}
where $R\sim 0.3\times 10^{24}$ sec.$^{-1}$, which yields 
$\tau_{Nuc.}\sim 
3\times 10^{32}$ yrs. 
Present experimental limits on $\tau_{Nucl.}$ from various nucleon decay 
searches e.g. at Kamiokande, Sudan, IMB, Super-Kamiokande, SNO are 
given below in table I with the value of $\tau_{n-\bar{n}}$ noted 
there\cite{nnbar}.
\begin{center}
\begin{tabular}{|c||c||c||c|}\hline\hline
Expt & source of neutrons & $\tau_{Nucl.}(yrs)$ & $\tau_{osc.}$(sec)\\ 
\hline 
Soudan & $^{56}Fe$ & $0.72\times 10^{32}$  & $1.3\times 10^8$\\ \hline
Frejus & $^{56}Fe$ & $0.65\times 10^{32}$  & $1.2\times 10^8$ \\ \hline
Kamiokande & $^{16}O$ & $0.43\times 10^{32}$ & $1.2 \times 10^8$ \\ 
\hline
Super-K &  $^{16}O$ & $1.77\times 10^{32}$ & $2.3\times 10^8$ \\ 
\hline\hline
\end{tabular}

Table Caption: Experimental lower limits on $\tau_{n-\bar{n}}$ and the 
nuclei used for deducing the limit.
\end{center}
One can also anticipate a similar bound 
from the SNO experiment\cite{bergevin}. 
This implies  $\rightarrow  \delta m \leq 
10^{-29}$ MeV. The nucleon decay 
searches for $\tau_{n-\bar{n}}$ become less efficient due to 
atmospheric neutrino induced background as we move to higher precision 
goals for  $\tau_{n-\bar{n}}$. It would 
therefore appear that in order to 
extend the  $\tau_{n-\bar{n}}$ search much beyond the current limit, one 
needs to focus more on free neutron oscillation experiments rather than than 
nucleon decay search.

We will see that present reactor neutron fluxes are precisely in
the right range to probe these values of $\tau_{n-\bar{n}}$ that is 
compatibitible with matter stability as well as latest limits on proton 
decay.
 \begin{figure}\begin{center}
\includegraphics[scale=0.8]{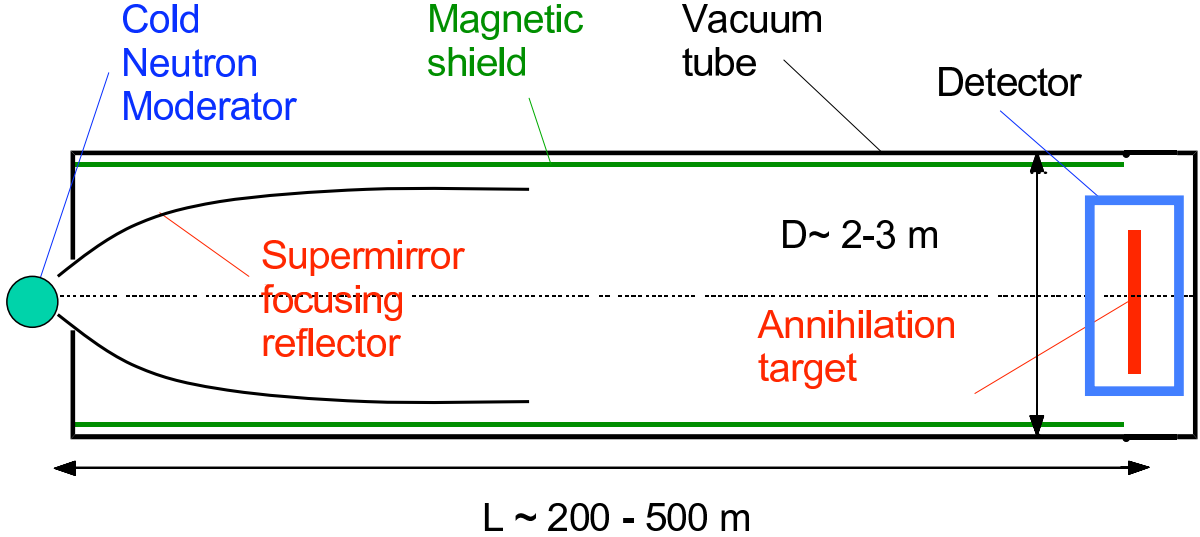}
\end{center}
\caption{ Typical horizontal reactor set-up for $n-\bar{n}$ oscillation 
search}. 
\end{figure}
\subsection{ Reactor Experiment to search for neutron oscillation}
The basic idea behind a reactor search for $n-\bar{n}$ oscillation is 
to have a cold neutron beam from a reactor (neutron speed $v\sim 1-2$ 
kilo meters/sec.) and have the beam pass through a high vacuum, demagnetized 
region for as long as feasible to a detector at the end of this 
``tunnel''. If one of the neutrons in the beam oscillates to an 
anti-neutron, this will annihilate in the detector and create a 
multi-pion signal with total energy of $\sim 2$ GeV. The need for 
demagnetization comes from the fact that the magnetic field of the 
Earth ($\sim 0.5$ Gauss) will split 
the neutron and anti-neutron energy levels by $\mu_n B\sim 3 \times 
10^{-21}$ GeV. If we take a typical flight time in the ``tunnel'' of 
about one sec., it will imply $\mu_n B t sim 3\times 10^{-21}$ GeV sec 
which about a 1000 times $\hbar$ and hence does not satisfy the condition 
for free neutron oscillation. It is therefore necessary to reduce the 
magnetic field to the level of $10^{-4}-10^{-5}$ Gauss. This is known 
technology that uses $\mu$-metal shielding. The typical horizontal set-up 
for such an experiment is given in Fig. 1 above. Under these conditions, 
one gets for the number of anti-neutrons for a given neutron beam the figure 
of merit of the experiment:
\begin{eqnarray}
\# of events~=~\Phi_n\left(\frac{t}{\tau_{N-\bar{N}}}\right)^2 \times
T 
\end{eqnarray}
where $\Phi_n$ is the reactor flux; $v_n t$= distance to detector; $T$
running time. Maximum reactor fluxes for a 100 MW reactor is about $\sim
10^{13}-10^{14}$ neutrons per sec.; for $t=0.1$ sec. and $T\sim 3$
years can yield a limit of $10^{10}$ sec. Using this technique, the first 
$n-\bar{n}$ search was carried out at the ILL laboratory in Grenoble, 
France and a lower limit of $\tau_{n-\bar{n}}\geq 
8.6\times 10^7$ sec. was obtained\cite{milla}.
New searches have recently been proposed which could improve this limit 
by two orders of magnitude\cite{kamyshkov}. Fig. 2 compares the free 
oscillation time to the nuclear instability life time (figure courtesy of 
Y. Kamyshkov Ref.\cite{kamyshkov}).
 \begin{figure} \begin{center}
\includegraphics[scale=0.8]{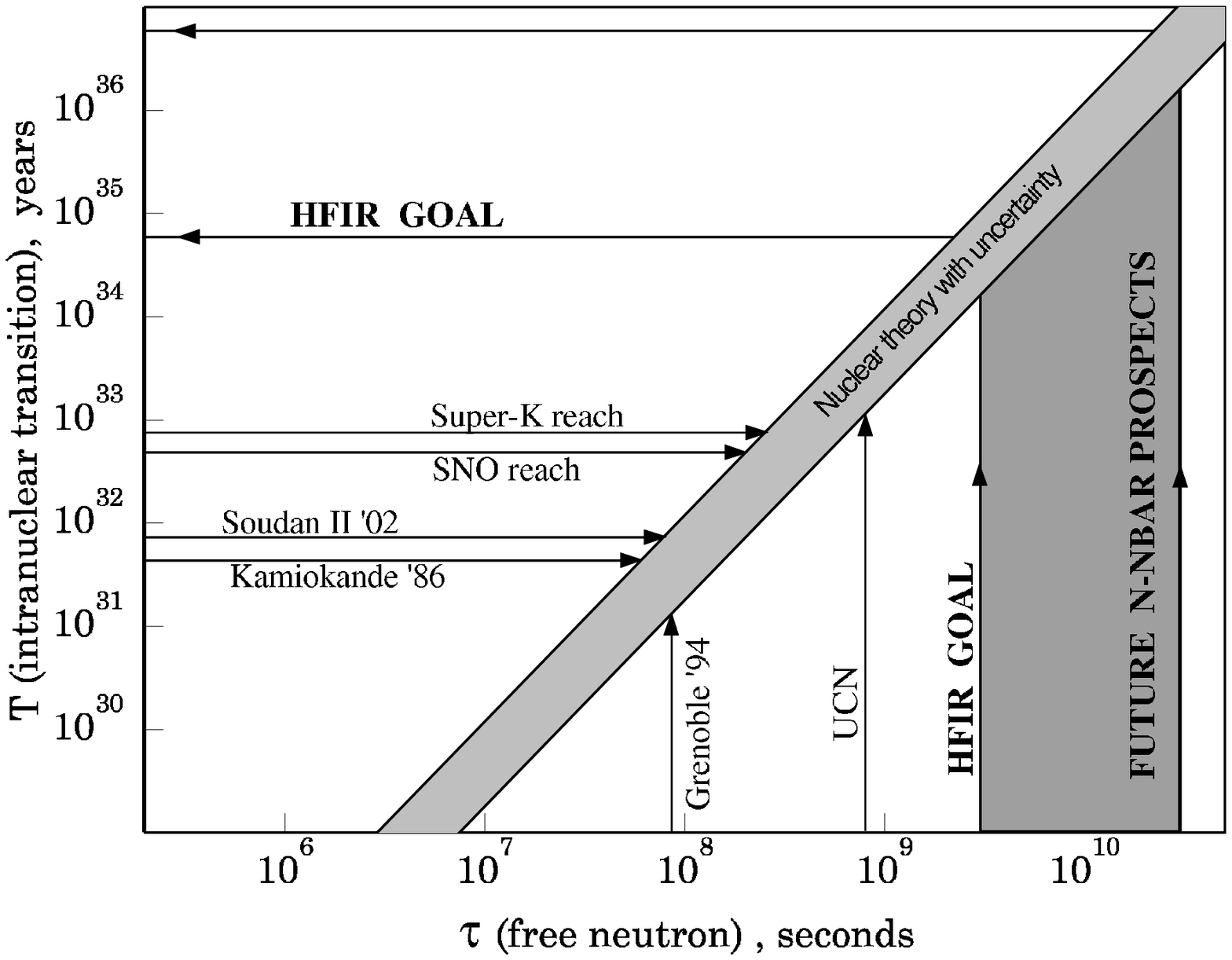} 
 \end{center}
\caption:{Comparision of free neutron oscillation time vs nuclear 
instability life time}
\end{figure} 
It is clear that time of flight $t$ anf the flux are two crucial factors 
that determine how precisely one can measure the $n-\bar{n}$ transition 
time. Clearly, the slower the neutrons the bigger is the time of flight. 
A possibility that is being pursued is to use ultracold neutrons with 
$\sim$ 100 nano-eV kinetic energies which move with speeds of order 4-5 
m/sec and are trapped in a bottle\cite{young} and may become competitive 
with cold reactor beam experiments in near future\cite{snow}.

 \section{Operator analysis of $\Delta B=2$ processes and new physics 
scale probed} Whenever a new process is used to probe physics beyond the 
standard model, a general question that is essential to know is the mass 
scale probed by it. A simple way to get a general idea about the scale 
probed by a 
physical process is to do an operator analysis i.e. assuming a particular
symmetry and spectrum of the low energy theory, write effective
higher dimensional operators for the process under consideration
and see for what value of the mass scaling the operator, the
process is observable. This argument has been a useful tool to
probe physics at short distance scales probed by processes such as proton 
decay, neutrino mass as well as corrections to standard model 
observables.

The method however has its limitations. For instance,
often in these discussions, one uses the SM fermion spectrum but
if there is a SM non-singlet new particle with a 100 GeV mass not
discovered yet, the naive scale arguments can be misleading since new 
operators may appear. 
Similarly, the presence of unknown higher symmetries could also
invalidate these arguments. 
We will apply this discussion to the $\Delta B=2$ operators below to see 
the scale probed by neutron-anti-neutron oscillation. 

\subsection{Standard model operator}
In the standard model, the effective operator responsible for $\Delta 
B=2$ and $\Delta L =~0$ processes is given by:
\begin{eqnarray}
{\bf O}_{\Delta B=2}~=\frac{1}{M^5}[QQQQ(d^cd^c)^*+ u^cd^cd^cu^cd^cd^c]
\end{eqnarray}
The strength of this operator $G_{\Delta B=2}\simeq\frac{1}{M^5}$. 
 The $n-\bar{n}$ mixing mass can be deduced from
this operator by simple dimensional analysis to be:
\begin{eqnarray} \delta m_{n-\bar{n}}\simeq c G_{\Delta 
B=2}\Lambda^6_{QCD}
=c\frac{\Lambda^6_{QCD}}{M^5}\end{eqnarray}
 where we chooze $c$ to be of order
one. Attempts have been made in the past to estimate $c$ using bag as 
well as other phenomenological models for hadrons\cite{bag}. Taking the 
best current lower bound on
$\tau_{n\leftrightarrow \bar{n}}$ from ILL reactor experiment
\cite{milla} which is $10^{8}$ sec. and comparable bounds from
nucleon decay search experiments \cite{nnbar}, one can then obtain
a lower limit on $M$ to be around 10-300 TeV depending on other 
couplings in a theory.
If proposals to improve the precision of this search by at
least two orders of magnitude \cite{kamyshkov} are carried out,
, then in the context of the standard model particles, the mass scale 
probed will be around a 1000 TeV. Thus we see that neutron-anti-neutron 
oscillation will probe physics at scales far below the GUT scales that 
proton decay will probe. An important question to ask is whether the 
scale probed can be higher in the presence of new physics around the TeV 
scale. We discuss this below.

 \subsection{Supersymmetry and enhancement of $\Delta B=2$ operator}
In the presence of supersymmetry at the TeV scale, there are new
particle such as squarks and sleptons with TeV scale mass. They
can then enter the effective $\Delta B=2$ operator. 
This kind of effect is familiar from discussions of supersymmetric 
grand unified theories where the presence of super-partners with TeV 
scale masses can drastically alter the operator 
analysis for nucleon decay. For instance the dimension six operator 
responsible for proton decay in non-susy models goes like $QQQL/M^2$ 
whereas the presence of squarks with TeV  or sub-TeV masses, the dominant 
operator becomes 
$QL\tilde{Q}\tilde{Q}/M$, which has reduced power dependence on mass 
and is well known to put severe constraints on grand unified theories.

A typical leading operator for $n-\bar{n}$ oscillation for supersymmetric 
case is:
\begin{eqnarray}
{\bf O}_{\Delta
B=2}~=~\frac{1}{M^3}u^cd^c\tilde{d^c}\tilde{u^c}\tilde{d^c}\tilde{d^c}
\end{eqnarray}
Note that the power dependence on the seesaw scale (or B-L
breaking scale) has now considerably softened. The conversion of susy 
particles to SM particles brings some suppression; but the the overall 
impact is that  simple power counting arguments change. They imply that 
if this is the leading operator
one can probe the seesaw scale upto $10^8$ GeV with $n-\bar{n}$ 
oscillation times of $10^{10}-10^{11}$ sec.
The power dependence in fact softens even further if there are new
color sextet particles at or below the TeV scale. For instance, if
there is a scalar diquark coupled sextet field of $\Delta_{u^cu^c}$ type,
the leading operator becomes:
\begin{eqnarray}
{\bf O}_{\Delta
B=2}~=~\frac{1}{M^2}d^cd^c\Delta_{u^cu^c}\tilde{d^c}\tilde{d^c}
\end{eqnarray}
and the scale reach of $n-\bar{n}$ goes up to $10^{11}$ GeV. If on
the other hand there is a field $\Delta_{u^cd^c}$, the leading
order operator becomes:
\begin{eqnarray}
{\bf O}_{\Delta
B=2}~=~\frac{1}{M}d^cd^c\Delta_{u^cd^c}\Delta_{u^cd^c}
\end{eqnarray}
increasing the scale reach to the GUT scale.
The question then arises whether there are plausible models where
this can happen. Below I give an example of models which have TeV scale 
$\Delta_{u^cu^c}$ fields naturally so that an $1/M^2$ dependence on the 
seesaw scale is quite natural. From this, we see that scale 
of observable $n-\bar{n}$ physics can be
anywhere from ~ 300 TeV to $10^13$ TeV

\section{Neutrino mass physics probed by $n-\bar{n}$ oscillation }
 To see what new physics is revealed by the search 
for $n-\bar{n}$ oscillation, we give several examples where low scale for 
seesaw  manifests in the non-leptonic sector of the theory as $n-\bar{n}$ 
oscillation. This happens due to the connection of seesaw to
 local or global B-L symmetry. There could also be other kind of beyond 
the standard model scenarioss where extra space 
dimensions curled up to TeV$^{-1}$ length scales and where additional 
dynamical inputs can lead to observable $n-\bar{n}$ oscillation. In the 
former case, it will throw important light on the physics associated 
with neutrino mass and we discuss this below. 

 There are various reasons to think that
 B-L symmetry is an intimate feature of physics beyond the
 standard model. Some of them are as follows:
\begin{itemize}

\item (i) The seesaw mechanism 
for understanding small
 neutrino masses requires the introduction of  right
 handed neutrinos\cite{seesaw}. In the presence of three right handed 
neutrinos it is more natural for
 B-L symmetry which was a global symmetry of the SM Lagrangian to 
become a local symmetry;

 \item (ii) An inherent aspect of seesaw is the Majorana mass
  of the right handed neutrino that breaks the B-L symmetry providing one 
way to understand why the seesaw scale is so much less than the Planck 
scale; 

\item (iii) A third reason appears once one admits the presence of 
supersymmetry
  at the TeV scale, as is widely believed and requires the
  lightest SUSY particle (LSP) to be naturally stable in order to be the 
dark
  matter candidate of the universe. The simplest way to have SUSY LSP 
naturally
  stable is to have B-L as a symmetry of physics beyond MSSM\cite{Rp}.

\end{itemize}
  
If indeed B-L symmetry is present in nature, there are several
  questions that immediately come to mind: 

\begin{itemize}

\item (a) Is it a global or local symmetry ? 

\item (b) is it a broken or unbroken 
symmetry ?

\item  (c)
  if it is broken, what is the breaking scale and what is the
  associated physics ? 

\end{itemize}

 I will now argue that search for
  the process of neutron-anti-neutron 
oscillation\cite{kuzmin,glashow,marshak} can provide a
  partial answers to some  of these important questions: in
  particular the question of scale of B-L breaking and associated 
expanded gauge symmetry and hence physics associated with the origin of 
neutrino mass. The point is that since a $\Delta B=2$ operator breaks B-L 
by two units, it is natural to associate the scale associated with the 
$\Delta B=2$ operator also with the scale of B-L breaking since neither 
$B$ nor $L$ are separate symmetries of the standard model but rather the 
symmetry is $B-L$ . Since seesaw mechanism also breaks B-L symmetry, $M$ 
could also be the seesaw scale. In fact, in the context of models such as 
those based on $SU(2)_L\times SU(2)_R \times SU(4)_c$\cite{ps}, the same 
mass scale $M$ is responsible for both processes as was first shown in 
Ref.\cite{marshak}. Another purely group theoretical way to see this is 
to note that in left-right symmetric models, there is a relation between 
the electric charge and the B-L generator as follows: 
\begin{eqnarray} 
Q~=~I_{3L}~+~I_{3R}~+\frac{B-L}{2}
\end{eqnarray}
For distance scales shorter than the electroweak scale, we have $\Delta 
I_{3L}=0$ and electric charge conservation then implies that $\Delta 
I_{3R}~=~-\frac{B-L}{2}$ implying that parity violation (or $\Delta 
I_{3R}~=~1$) implies not only that $\Delta L~=~2$ i.e. seesaw mechanism 
for neutrino mass but in the non-leptonic sector it implies $\Delta 
B~=~2$ which is of course $n-\bar{n}$ oscillation.

 Thus the 
search for $n-\bar{n}$
oscillation could not only illuminate the nature of the important
symmetry, B-L but could also be one way to unravel the mystery of
the seesaw mechanism that is expected to be major player in the
physics of neutrino mass as well as the origin of parity violation in 
Nature.

\section{$SU(2)_L\times SU(2)_R\times SU(4)_c$ model with light
diquarks} In this section we demonstrate by an explicit example how 
neutrino mass and $n-\bar{n}$ are intimately connected in this partial 
unification theory. We first recapitulate some elementary facts about the 
model.

 The quarks and leptons in this model are unified and 
transform as $\psi:({\bf 2,1,4})\oplus \psi^c:({\bf 1,2},\bar{\bf 4})$
representations of $SU(2)_L\times SU(2)_R\times SU(4)_c$. For the
Higgs sector, we choose, $\phi_1:(\bf{2,2,1})$ and
$\phi_{15}:(\bf{2,2,15})$ to give mass to the fermions. The
$\Delta^c:({\bf 1,3,10})\oplus \bar{\Delta}^c:({\bf
1,3},\overline{\bf 10})$ to break the $B-L$ symmetry. There are 
color sextet scalar diquark
fields contained in the $\Delta^c:(\bf{1,3,10})$ multiplet, which will 
play a crucial role in generating $n-\bar{n}$ oscillation. We first 
discuss the non-0supersymmetric version of the model in next subsection 
and follow it up by the supersymmetric generalization where the scale 
reach of $n-\bar{n}$ goes up all the way to $10^{10}$ GeV or so.

\subsection{Non-supersymmetric version}
The Higgs potential for this model can be written as :
\begin{eqnarray}
V(\phi_0,\phi_{15}, \Delta_{R,L})~=~V_0~+~V_m
\end{eqnarray}
where 
\begin{eqnarray}
V_m~=~\lambda_m 
\epsilon^{\mu\nu\sigma\rho}\epsilon^{\mu'\nu'\sigma'\rho'}
\Delta_{R,\mu\mu'}\Delta_{\nu\nu'}\Delta_{\sigma\sigma'}\Delta_{\rho\rho'}+~ 
(R\leftrightarrow L)+~h.c. 
\end{eqnarray}
This term is crucial for generating baryon number non-conservation in 
this theory and without it, the model cannot explain the origin of 
matter in the universe (as we see below). We will see that the same term 
is responsible for the $\Delta B=2$ $n-\bar{n}$ oscillation.

In order to discuss neutrino masses and $n-\bar{n}$ oscillation, let us 
write down the Yukawa couplings in the model:
\begin{eqnarray}
{ L}_Y~=~h_1\bar{\Psi}_L\Phi_1 \Psi_R~+~h_{15}\bar{\Psi}_L\Phi_{15} 
\Psi_R~+~f\Psi^T_RC^{-1}\Delta_R\Psi_R~+(R\leftrightarrow L)+H.C.
\end{eqnarray}
Using standard spontaneous breaking via the vev of $(1,3,10)$ to the 
standard model gauge group and $\Phi_{1,15}$ to give mass to both to 
charged fermions and Dirac mass to neutrinos, we implement the seesaw 
mechanism for neutrinos. It was shown in ref.\cite{marshak} that the same 
$(1,3,10)$ via the potential term $V_m$ leads to $n-\bar{n}$ oscillation 
via the diagram in Fig. 3. 
\begin{figure}[h!]\begin{center} 
\psfig{file=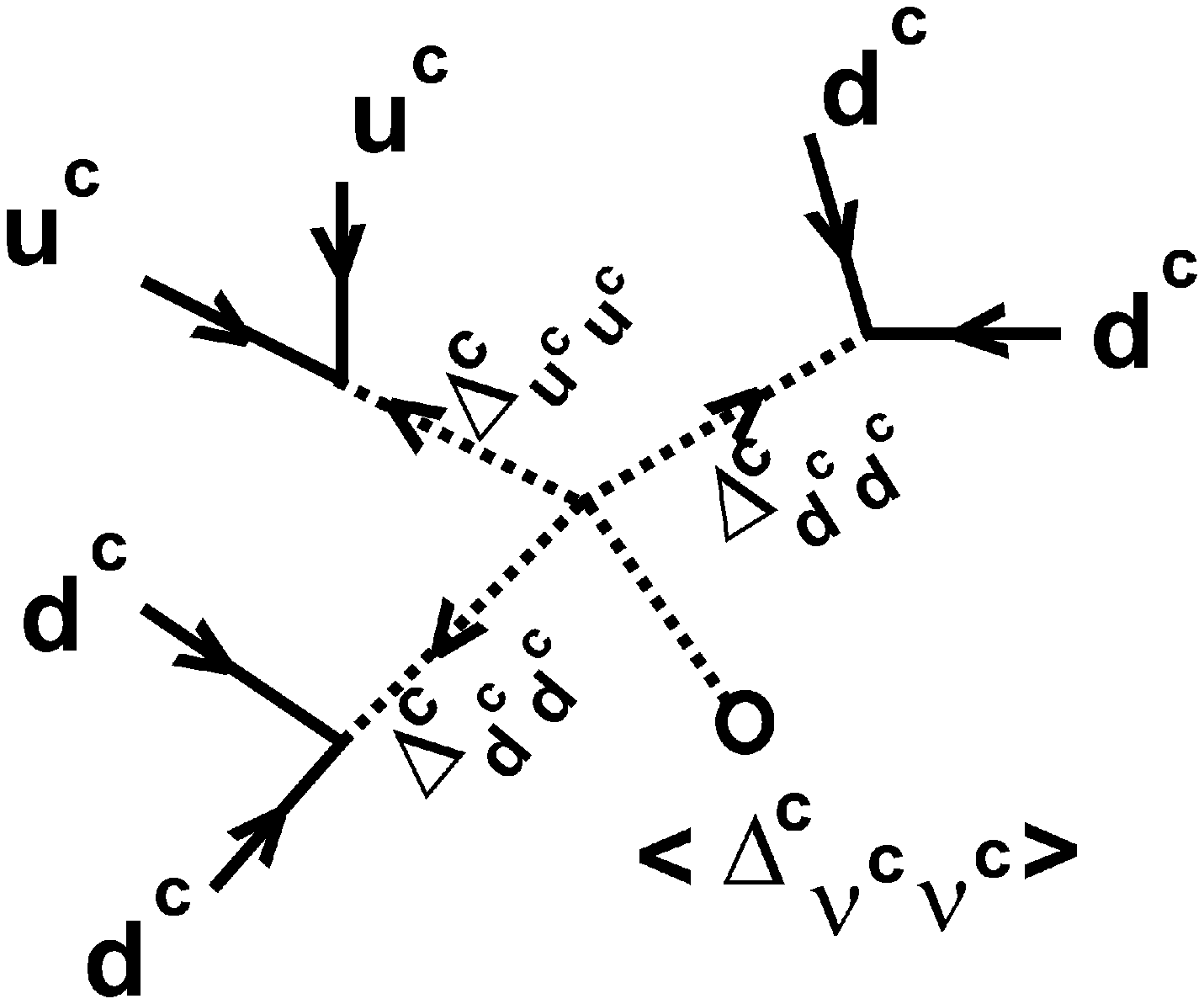,width=4.5in} \caption{A Feynman
diagram contributing to $n-\bar{n}$ oscillation as discussed in
Ref. 8. Another class of graphs involves two $\Delta_{u^cd^c}$ and one 
$\Delta_{d^cd^c}$ scalar exchanges.}\end{center}
\end{figure}
The strength of $n-\bar{n}$ oscillation is given by:
\begin{eqnarray}
G_{\Delta B=2}~=~\frac{\lambda f^3_{11}v_{BL}}{M^6_{\Delta_{qq}}}
\end{eqnarray}
We see the fifth power dependence on the seesaw scale since the diquarks 
are expected to be at seesaw scale by the usual naturalness 
considerations. Also note that even after symmetry breaking this model 
has a $Z_2$ symmetry given by $(-1)^B$.
  
It has recently been shown\cite{bdm} that this model also explains the 
baryon asymmetry of the universe via the mechanism of post-sphaleron 
baryogenesis to be discussed below\cite{babu1}.

\subsection{Supersymmetric version}
We now discuss the supersymmetric version of this model where the triplet 
fields doubled to cancel the gauge anomalies. We also add a
$B-L$ neutral triplet $\Omega:(\bf{1,3,1})$ which helps to reduce the 
accidental global symmetry of the model and hence the number
 of light diquark states.
 The superpotential of this model is given
by:
\begin{eqnarray}
W~=~W_Y~+~W_{H1}+~W_{H2}+~W_{H3}
\end{eqnarray}
where
\begin{eqnarray}
W_{H1}&=&\lambda_1 S( \Delta^c\bar{\Delta}^c-M^2)~+\lambda_C
\Delta^c\bar{\Delta}^c\Omega~ +\mu_{i}{\rm Tr}\,(\phi_i\phi_i)
\end{eqnarray}
\begin{eqnarray}
W_{H2}~=~\lambda_A
\frac{(\Delta^c\bar{\Delta}^c)^2}{M_{P\!\ell}}
+ \lambda_B\frac{(\Delta^c{\Delta^c})
(\bar{\Delta}^c\bar{\Delta}^c)}{M_{P\!\ell}}
\end{eqnarray}
\begin{eqnarray}
 W_{H3}~=~ \lambda_D \frac{{\rm Tr}\,(\phi_1\Delta^c
\bar{\Delta}^c\phi_{15})}{M_{P\!\ell}} \,, \\
\end{eqnarray}
\begin{eqnarray}
W_Y&=&h_1\psi\phi_1 \psi^c + h_{15} \psi\phi_{15} \psi^c + f
\psi^c\Delta^c \psi^c.
\end{eqnarray}
Note that since we do not have parity symmetry in the model, the
Yukawa couplings $h_1$ and $h_{15}$ are not symmetric matrices.
When $\lambda_B=0$, this superpotential has an accidental global
symmetry much larger than the gauge group\cite{chacko}; as a
result, vacuum breaking of the $B-L$ symmetry leads to the
existence of light diquark states that mediate $N\leftrightarrow
\bar{N}$ oscillation and enhance the amplitude. In fact it was
shown that for $\langle\Delta^c\rangle\sim
\langle\bar{\Delta}^c\rangle\neq 0$ and $\langle\Omega\rangle\neq
0$ and all VEVs in the range of $10^{11}-10^{12}$ GeV, the light
states are those with quantum numbers: $\Delta_{u^cu^c}$.
 The symmetry argument behind is that \cite{chacko}
 for $\lambda_B=0$, the above superpotential is invariant under
$U(10,c)\times SU(2,c)$ symmetry  which breaks down to
$U(9,c)\times U(1)$ when $\langle\Delta^c_{\nu^c\nu^c}\rangle =
v_{BL}\neq 0$. This results in 21 complex massless states; on the
other hand these vevs also breaks the gauge symmetry down from
$SU(2)_R\times SU(4)_c$ to $SU(3)_c\times U(1)_Y$. This allows
nine of the above states to pick up masses of order $gv_{BL}$
leaving 12 massless complex states which are the six
$\Delta^c_{u^cu^c}$ plus six $\bar{\Delta}^c_{u^cu^c}$ states.
Once $\lambda_B\neq 0$ and is of order $10^{-2}-10^{-3}$, they
pick up mass (call $M_{u^cu^c}$) of order of the elctroweak scale.

\subsection{A new diagram for neutron-anti-neutron oscillation}
To discuss $n\leftrightarrow \bar{n}$ oscillation,
we introduce a new term in the superpotential of the form\cite{marshak}:
\begin{eqnarray}
W_{\Delta B=2}~=~\frac{1}{M_*}\epsilon^{\mu'\nu'\lambda'\sigma'}
\epsilon^{\mu\nu\lambda\sigma}
\Delta^c_{\mu\mu'}\Delta^c_{\nu\nu'}\Delta^c_{\lambda\lambda'}
\Delta^c_{\sigma\sigma'}\,, \label{Delta_B=2}
\end{eqnarray}
where the $\mu,\nu  $ etc stand for $SU(4)_c$ indices and we have
suppressed the $SU(2)_R$ indices. Apriori $M_*$ could be of order
$M_{P\ell}$; however the terms in Eq.(2) are different from those
in Eq. (4); so they could arise from different a high scale
theory. The mass $M_*$ is therefore a free parameter that we
choose to be much less than the $M_{P\ell}$. This term does not
affect the masses of the Higgs fields. When
$\Delta_{\nu^c\nu^c}^c$ acquires a VEV, $\Delta B = 2$ interaction
are induced from this superpotential,
%
and $n\leftrightarrow \bar{n}$ oscillation are generated by two
diagrams given in Fig. 3 and 4. The first diagram (Fig. 3) in
which only diquark Higgs fields are involved was already discussed
in  \cite{marshak} and goes like $G_{n\leftrightarrow \bar
n}\simeq \frac{f^3_{11}v_{BL}
M_\Delta}{M^2_{u^cu^c}M^4_{d^cd^c}M_*}$,
 Taking $M_{u^cu^c}\sim 350$ GeV, $M_{d^cd^c}\sim \lambda'v_{BL}$ and
$M_\Delta \sim v_{BL}$ as in the argument \cite{chacko}, we see
 that this
diagram scales like $v^{-3}_{BL} v_{wk}^{-2}$.
In ref.\cite{DMM} a new diagram (Fig. 4) was pointed out which owes its
origin to supersymmetry. We get for its contribution to $G_{\Delta
B=2}$:
\begin{eqnarray}
G_{n \leftrightarrow \bar n}\simeq\frac{g^2_3}{16\pi^2}
\frac{f^3_{11}v_{BL}}{M^2_{u^cu^c}M^2_{d^cd^c}M_{\rm SUSY}M_*}.
\end{eqnarray}
Using the same arguments as above, we find that this diagram
scales like $v^{-2}_{BL} v_{wk}^{-3}$ which is therefore a significant
enhancement over diagram in Fig.3 when $v_{BL}\gg v_{wk}$.
\begin{figure}[h!]\begin{center}
\psfig{file=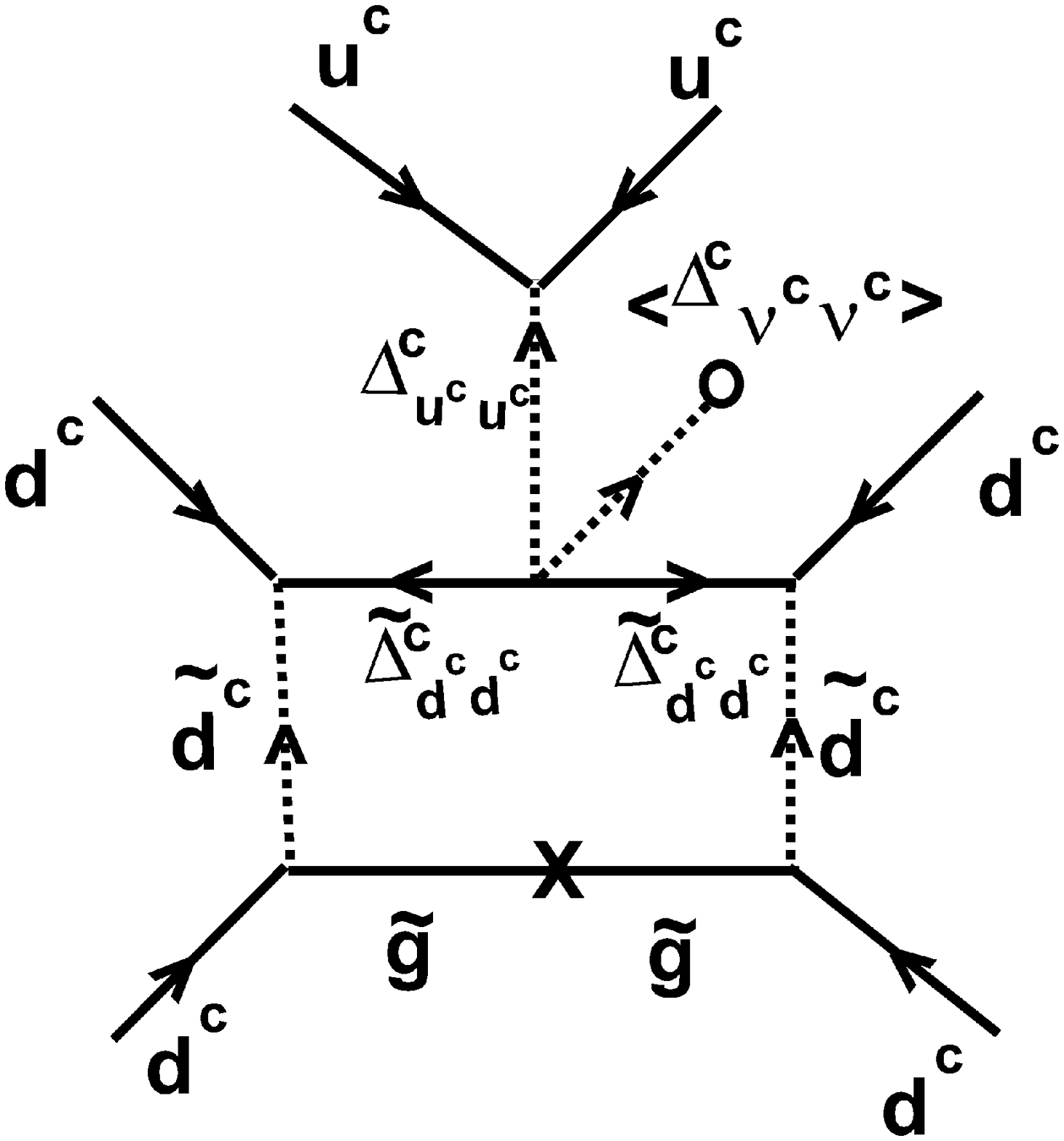,width=4.5in} \caption{The new Feynman
diagram for $n-\bar{n}$ oscillation.}
\end{center}
\end{figure}
In order to estimate the rate for $n\leftrightarrow \bar{n}$
oscillation, we need not only the different mass values for which
we now have an order of magnitude, we also need the Yukawa
coupling $f_{11}$. Now $f_{11}$ is a small number since its value
is associated with the lightest right-handed neutrino mass.
However, in the calculation we need its value in the basis where
quark masses are diagonal. We note that the
 $n-\bar{n}$ diagrams involve only the
right-handed quarks, the rotation matrix need not be the CKM
matrix. The right-handed rotations need to be large e.g. in order
to involve $f_{33}$ (which is $O(1)$), we need
$(V_R^{(u,d)})_{31}$ to be large, where $V_L^{(u,d)\dagger}
Y_{u,d}V_R^{(u,d)}=Y_{u,d}^{\rm diag.}$. The left-handed rotation
matrices $V_L^{(u,d)}$ contribute to the CKM matrix, but
right-handed rotation matrices $V_R^{(u,d)}$ are unphysical in the
standard model. In this model, however, we get to see its
contribution since we have a left-right gauge symmetry.
Let us now estimate the time of oscillation. When we start with a
$f$-diagonal basis (call the diagonal matrix $\hat f$), the
Majorana coupling $f_{11}$ in the diagonal basis of up- and
down-type quark matrices can be written as $(V_R^T {\hat f}
V_R)_{11} \sim (V^R_{31})^2 \hat f_{33}$. Now $\hat f_{33}$ is
$O(1)$ and $V^R_{31}$ can be $\sim 0.6$, so $f_{11}$ is about 0.4
in the diagonal basis of the quark matrices. We use $M_{\rm
SUSY}$, $M_{u^cu^c}$ $\sim 350$ GeV and $v_{BL}\sim 10^{12}$ GeV.
 The mass of $\tilde\Delta_{d^cd^c}$ i.e. $M_{
d^cd^c}$ is $10^{9}$ GeV which is obtained from the VEV of
$\Omega:(\bf{1,3,1})$. We choose $M_*\sim 10^{13}$ GeV. Putting
all the above the numbers together, we get
\begin{eqnarray}
G_{n \leftrightarrow \bar n}\simeq 1 \cdot 10^{-30} \ {\rm
GeV^{-5}}. \label{G-NNbar}
\end{eqnarray}
Taking into account the hadronic matrix element effect, the
$n-\bar{n}$ oscillation time is found to be about $2.5\times
10^{10}$ sec which is within the reach of possible next generation
measurements. If we chose, $M_*\simeq M_{P\ell}$, we will get for
$\tau_{n-\bar{n}}\sim 10^{15}$ sec. unless we choose the
$M_{d^dd^c}$ to be lower (say $10^7$ GeV). This is a considerable
enhancement over the nonsupersymmetric model of \cite{marshak}
with seesaw scale of $10^{12}$ GeV.
We also note that as noted in \cite{marshak} the model is
invariant under the hidden discrete symmetry under which a field
$X \rightarrow e^{i\pi B_{X}}X$, where $B_X$ is the baryon number
of the field $X$. As a result, proton is absolutely stable in the
model. Furthermore, since R-parity is an automatic symmetry of
MSSM, this model has a naturally stable dark matter.

\section{Baryogenesis and $n-\bar{n}$ oscillation}
In the early 1980's when the idea of neutron-anti-neutron
oscillation was first proposed in the context of unified gauge
theories, it was thought that the high dimensionality of the
$\Delta B\neq 0$ operator would pose a major difficulty in
understanding the origin of matter in the Universe. The main reason for 
this assessment 
is that the higher dimensional operators remain in thermal
equilibrium until late in the evolution of the universe without 
contradicting any low energy observations. This is because the
thermal decoupling temperature $T_*$ for such interactions goes roughly
like $v_{BL}\left(\frac{v_{BL}}{M_{P}}\right)^{1/9}$ which can be in the
range of temperatures where B+L violating sphaleron transitions are 
in full thermal equilibrium, say for example $v_{BL}\simeq 10-100$ TeV, 
as required for the case of observable $n-\bar{n}$ oscillation. They 
will therefore erase any baryon
asymmetry generated in the very early moments of the universe
(say close to the GUT time of $10^{-36}$ sec. or so) in then prevalent
baryogenesis models. Even though GUT baryogenesis models are no more 
popular, the same argument will apply to any other high temperature 
baryogenesis mechanism. 

In models
with observable $n-\bar{n}$ oscillation therefore, one has to
search for new mechanisms for generating baryons below the weak
scale. In this section, we discuss such a possibility\cite{babu1} which 
was discussed in a recent paper. As we see below, it is ideally suited 
for embedding into the $G_{224}$ model discussed in Ref.\cite{marshak} 
and leads to an interesting side ``bonus'' that it puts an upper limit on 
the neutron-anti-neutron oscillation time $\tau_{n-\bar{n}}$. 

As an illustration of the way the new mechanism operates, let us assume
that there is a complex scalar field that couples to the $\Delta B=2$ 
operator as follows
i.e. $L_I~=~Su^cd^cd^cu^cd^cd^c/M^6$, where the scalar boson has mass of
order of the weak scale and $B=2$. When $<S>\neq 0$, this interaction 
leads to baryon number violation by two units and observable $n-\bar{n}$ 
transition if $M$ 
is in the few hundred to 1000 GeV range. Writing 
$S~=~\frac{1}{\sqrt{2}}(v_{BL}~+~S_r+S_I)$, we see that the direct decay 
of $S_r$ involves both $\Delta B=\pm 2$ final states and thereby 
satisfies the first requirement for baryogenesis.

The first point to note is that the high
dimension of $L_I$ allows the scalar $\Delta B\neq 0$ decay to go out of 
equilibrium at weak scale temperatures. This satisfies the out of equilibrium
condition given by Sakharov for generating 
matter-anti-matter asymmetry. To see this, let us give some examples:
if we chose the proton decay operators such as $QQQL$, the decoupling 
temperature consistent with present experimental bounds on proton 
life time would be around $10^{15}$ GeV or so. So to apply our mechanism, 
we need to consider higher dimensional operators. A typical example of 
such an operator that we will focus on in this paper is of type: 
$u^cd^cd^cu^cd^cd^c/M^5$. As note earlier, this operator leads to 
$n-\bar{n}$ 
oscillation and for this to be in the observable range, the associated 
mass scale has to be in the few TeV range. The decupling temperature can 
then be easily estimated from the formula $T_d\sim 
M\left(\frac{10M}{M_P}\right)^{1/9}\sim 10^{-1.5}M$. For appropriate 
values of $M$ in the range of interest $T_d$ can be below 100 GeV so 
that the sphalerons have gone out of equilibrium and baryogenesis 
follows. To make these ideas more concrete, below we give an explicit 
example\cite{babu1}.

We consider an effective sub-TeV scale model that gives rise to the
above higher dimensional
operator  for ${n\leftrightarrow \bar{n}}$ oscillation. It consists of
the following color sextets, $SU(2)_L$
singlet scalar bosons $(X,Y,Z)$ with hypercharge $-\frac{4}{3},
+\frac{8}{3}, +\frac{2}{3}$ respectively that couple to quarks.
These states emerge naturally if this low scale theory is embedded into 
 the $SU(2)_L\times SU(2)_R\times SU(4)_c$ (denoted by $G_{224}$) 
model described in the previous section. The scalar field $S$ can be 
identified with the complex field $\Delta_{\nu^c\nu^c}$ whose vev breaks 
B-L symmetry by two units as is the case for us and $S_r\equiv 
Re(\Delta^0_{\nu^c\nu^c}$\cite{bdm}. 
The $X,Y,Z$ fields can be identified with the color sextet partners of 
$\Delta_{\nu^c\nu^c}$ in the $(1,3,10)$ multiplet i.e. the 
$\Delta_{d^cd^c}, \Delta_{u^cu^c}$ 
and $\Delta_{u^cd^c}$ respectively. In the $X,Y,Z$ notation, 
one can write down the following standard model invariant
interaction Lagrangian:
\begin{eqnarray}
{ L}_I~&=&~ h_{ij}X d^c_id^c_j + f_{ij} Yu^c_iu^c_j~+
\\\nonumber && g_{ij} Z (u^c_id^c_j+u^c_jd^c_j)  +~\lambda_1 S
X^2Y~+~\lambda_2 SXZ^2
\end{eqnarray}
The complex scalar field $S$ clearly has $B-L~=~2$. To see the 
constraints on
the parameters of the theory, we note
that the present limits on $\tau_{n-\bar{n}}\geq 10^8$ sec.
implies that the strength $G_{n-\bar{n}}$ of the the $\Delta B= 2$
transition is $\leq 10^{-28}$ GeV$^{-5}$. From the above figure, we 
conclude that
\begin{eqnarray}
G_{n-\bar{n}}\simeq \frac{\lambda_1 M_1 h^2_{11}
f_{11}}{M^2_YM^4_X}~+~\frac{\lambda_2 M_1h_{11}g^2_{11}}{M^2_XM^2_Z}\leq
10^{-28} GeV^{-5}.
\end{eqnarray}
For $\lambda_{1,2} \sim h_{11}\sim f_{11}\sim g_{11}\sim 10^{-3}$, we 
require $M_{1}\sim
M_{X,Y,Z}\simeq 1$ TeV to satisfy this experimental bound. In our 
discussion of generic models of this type, we will stay close to
this range of parameters and see how one can understand the baryon
asymmetry of the universe. The singlet field will play a key role in the
generation of baryon
asymmetry. We assume that $<S>\gg M_{X}$ and $M_{S_r}\sim 100-500$
GeV. It can then decay into final states with $B=\pm 2$ i.e. six quarks 
and six anti-quarks.

Note that the Lagrangian of Eq. (26), leads to tree level contribution to 
flavor changing neutral current processes such as $K-\bar K$, $B-\bar B$ 
mixings etc and that will restrict the form of the flavor structure of 
the coupling matrices suitably. These constraints can also be satisfied 
in a realistic $G_{224}$ embedded model which also explain the 
neutrino masses and mixing\cite{bdm}. This is an important consideration 
since the magnitude of the baryon asymmetry depends on this detailed 
flavor structure.

On the way to calculating the baryon asymmetry, let us first discuss
the out of equilibrium condition. As the temperature of the universe
falls below the masses of the $X,Y,Z$ particles, the annihilation
processes $X\bar{X}\rightarrow d^c\bar{d^c}$ (and analogous
processes for $Y$ and $Z$) remain in equilibrium. As a result, the
number density of $X,Y,Z$ particles gets depleted as $e^{-M_X/T}$ and 
only the $S_r$
particle survives along with the usual standard model particles. One
of the primary generic decay modes of $S_r$ is $S_r\rightarrow
u^cd^cd^cu^cd^cd^c$. There could be other decay modes which depend on
the details of the model and one has to ensure that the rates to other 
modes are smaller than the six quark mode.

The generic chain of events leading to baryogenesis in this model is the 
following. At $T\sim 
M_{S_r}$, the decay rate is smaller than the Hubble expansion rate.
As the Universe cools below this temperature, the decay rate remains  
constant whereas the
expansion rate of the universe is slowing down. So at a temperature
$T_d$ far below $M_S$, $S$ will start to decay when the decay rate 
$\Gamma\sim H$ with $T_d$ is given by:
\begin{equation}T_d \simeq (\frac{M_{P\ell}M_S^{13}}{(2\pi)^9M_X^{12}})^{1/2}
\end{equation}
Since the corresponding epoch must be above that the QCD phase 
transition temperature, this puts a constraint on the parameters of the
model. Typically, we need to have  $M_S/M_X\sim 0.5 $ or so due to high 
power 
dependence of $T_d$ on this ratio. This implies that the $X,Y,Z$ masses 
cannot be
arbitrarily high, since the heavier these particles are, the lower $T_d$
will be. We expect this upper limit to be in the TeV range at most as 
we show below .

It is well known that baryon asymmetry can arise only via the
interference of a tree diagram with a one loop diagram. The tree diagram
is clearly the one where $S\rightarrow 6 q$. There are however two 
classes
of loop diagrams that can contribute: one where the loop involves the 
same
fields $X,Y$ and $Z$. A second one involves W-exchange, which involves
only standard model
physics at this scale (Fig. 6). We find that the second contribution can
actually dominate. In fact, in the $G_{224}$ embedding of the model, the 
first diagram vanishes. The second diagram also has the advantage that it 
involves less number of arbitrary parameters and the source of CP 
violation in this case is the same as the CKM CP violation present in 
the standard model. The
observed baryon asymmetry is related to the primordial baryon asymmetry 
$\epsilon_B$ as $\eta_B\simeq 
(\frac{g^*_{rec}}{g^*_{T_d}})^{3/4}\epsilon_B)$ where
\begin{eqnarray}
\epsilon_B\simeq \frac{n_S}{n_\gamma}\frac{\Gamma(S\rightarrow
6q)-\Gamma(S\rightarrow 6\bar{q})}{\Gamma(S)}
\end{eqnarray}
\begin{figure}[htbp]
\epsfxsize=8cm
\centerline{\epsfbox{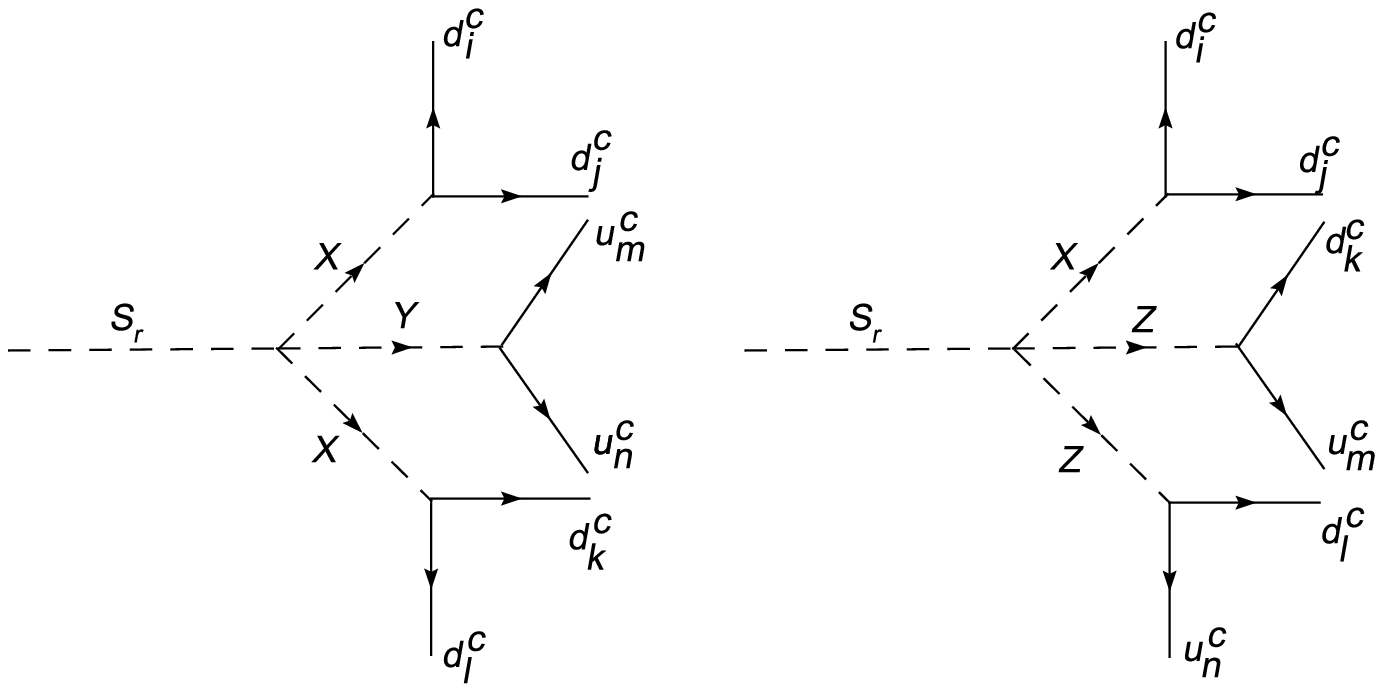}}
\caption{Tree level Feynman diagram for S decays.}
\end{figure}
\begin{figure}[h!]
\begin{center}
\epsfxsize=8cm
\includegraphics[scale=0.8]{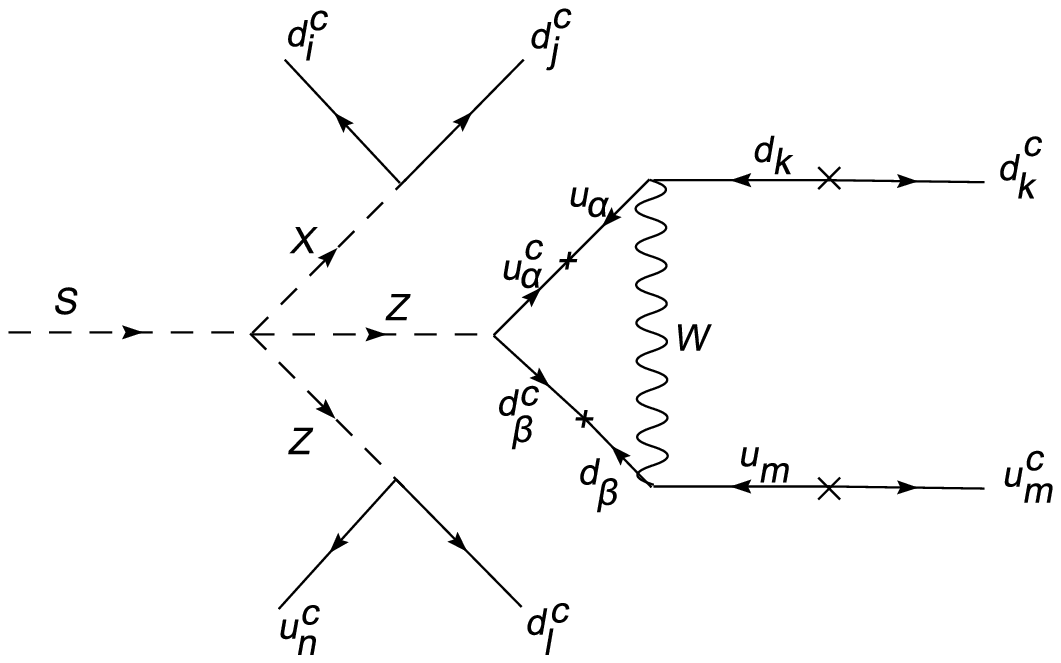}
\caption{One loop diagram for S decays.}
\end{center}
\end{figure}
We find that
\begin{equation}
{\epsilon_B^{\rm vertex} \over {\rm Br}} \simeq - {\alpha_2 \over
4} {{\rm ImTr}[gg^T \hat{M}_u V M_d g^\dagger M_u V^* \hat{M}_d] \over
{\rm Tr}(g^\dagger g) M^2_WM^2_S }~.
\label{asym}
\end{equation}
The magnitude of the asymmetry depends on $M_S$ as well as the detailed
profile of the various coupling matrices $h,g,f$ and we can easily
get the desired value of the baryon asymmetry by appropriately
choosing them. An important consequence of the above equation regardless 
of the details is that if $M_S$ is much bigger than about $500$ GeV, 
baryon asymmetry becomes very small. This implies that $M_{X,Y,Z}$ in 
turn should not be much larger than a TeV, implying that they can be 
accessible at the LHC. This also implies that the strength of $n-\bar{n}$ 
transition has a lower limit or an upper limit on $\tau_{n-\bar{n}}$.

There is a dilution of the baryon asymmetry arising from the
fact that $T_d \ll M_S$ since the decay of $S$ also releases entropy
into the universe.  Thius dilution factor is
\begin{equation}
d_B \simeq 
\epsilon_B\frac{(0.32g_*(T_d)T_d)^{3/4}}{(0.12M_S+0.32G_*T_D)^{3/4}}
\end{equation}
Since the decay rate of the $S$ boson depends inversely as a high power 
of $M_{X,Y}$, higher $X,Y$ bosons would imply that $\Gamma_S \sim H$ is
satisfied at a lower temperature and hence give a lower $d_B$. For our 
preferred choice of parameters i.e. $M_X\sim 2M_S \sim $ TeV, we find 
$d_B\sim 20\%$. Also for
choice of the coupling parameters $\lambda\sim f\sim h \sim g \sim
10^{-3}$, and $M_S \simeq 200\; GeV$ we find $\tau_{n-\bar{n}}\leq
10^{10}\;sec $.

We also note that if the model is embedded into a $G_{224}$ group, all 
three coupling matrices $h,g,f$ become equal; there are also some 
constraints on the gauge boson spectrum in this model\cite{bdm}. As noted 
above, an 
interesting consequence of adequate baryogenesis is that there are new 
sub-TeV- TeV scale color sextet scalar bosons which can be observed at 
LHC\cite{okada}. 

\section{Extra dimensional models with observable $n-\bar{n}$ 
oscillation} The possibility that there may be extra compact space 
dimensions has been under extensive theoretical as well as experimental 
investigation inspired by the fact that superstring theories do predict 
the existence of such models. In a large class of interesting models, 
these dimensions are believed to be of order TeV$^{-1}$ 
size\cite{dvali,raman} making 
their manifestations in many physical situations of experimental interest 
not only for colliders but also for low energy experiments. One class of 
such possibilities includes the breakdown of baryon number by two units.
We note two such discussions:

(i) In ref.\cite{nussinov}, it was noted that if there are millimeter 
size extra dimensions, one way to understand the observed fermion mass 
hierarchies would be to distribute the standard model fermions in the 
bulk. In such a scenario, their Yukawa couplings would be suppressed if 
they are further apart\cite{martin}. Thus in the context of a 5- or 6 
dimensional scenario, observed mass hierarchies pretty much fix the 
locations of the various standard model fermions in the bulk. One can 
then ask the question as to what happens to higher dimensional operators 
such as those giving proton decay and $n-\bar{n}$ oscillation. Since 
these are low scale models, one has to separate the quarks and leptons 
sufficiently to suppress these operators since their strength is given by 
the overlap integral of the various wave functions in the extra 
dimension. The $n-\bar{n}$ operator however involves only the right 
handed quarks and their locations is generally fixed by the mass 
hierarchy considerations. So there is no freedom to suppress their 
strengths. The strength of such operators was calculated in the context 
of six dimensional models in Ref.\cite{nussinov} and found to be in the
observable range for extra dimension sizes of order $45-100$ TeV, which 
are of the same order as being contemplated for phenomenological viable 
models of this kind. These considerations could also be applied to the 
case Randall-Sundrum models where the SM fermions are also generally 
expected to be distributed in the bulk.

(ii) A second class of models were discussed in ref.\cite{dvali1}, where 
general considerations of global charge separation from the branes is 
used to argue that one cannot have proton decay but observable 
$n-\bar{n}$ oscillations due to the fact that neutrons are both color and 
electric charge neutral. Such baryon number nonconservation occurs due to 
brane fluctuations which can create baby branes that carry baryon 
number if they are color and charge neutral. 
 
\section{R-parity breaking models} Finally, there have been explorations 
 of $n-\bar{n}$ oscilation to supersymmetry. As is well known, in minimal 
supersymmetric extensions of the standard model (MSSM), a baryon number
violating operator of the form $\lambda^{''}u^cd^cd^c$ is allowed in the 
superpotential. By itself it can lead to $\Delta B~=~2$ operator of the 
form $(u^cd^cd^c)^2$ at tree level via gluino exchange (Figure 7). 
Typical strength 
of such operators arising from this new interaction is given\cite{rpv} by:
\begin{eqnarray}
G_{\Delta B=2}~=~\frac{4\pi \alpha_s \lambda^{'' 
2}_{112}}{M^4_{\tilde{q}}M_{\tilde{G}}}
\end{eqnarray}
Note that due to color anti-symmetry, at the tree level, this interaction 
leads  $\Lambda-\bar{\Lambda}$ oscillation rather than $n-\bar{n}$
oscillation. However once radiative corrections are included, this 
operator via flavor changing effects gives rise to neutron 
oscillation\cite{sher}. This puts constraints on the R-parity violating 
couplings of the $\lambda^{''}$ type. It must be noted that if these 
couplings are present in combination with R-P violating operators of type
$QLd^c$, then it leads to catastrophic proton decay rates\cite{smirnov}; 
so in discussing $\Delta B~=~2$ transitions, we are assuming that $QLd^c$ 
operators are absent.
\begin{figure}
\begin{center}
\includegraphics[scale=0.8]{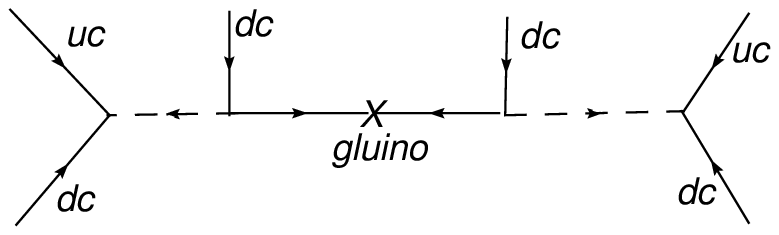} 
\caption{neutron oscillation induced in R-parity breaking MSSM}
\end{center} 
\end{figure} 

\section{Conclusion}
In summary, there is now a widely held belief that neutrino mass provides 
a clue to the existence of a new local B-L symmetry of 
nature. A key question in particle theory beyond the standard model
is then to find out the scale and detailed nature of physics associated 
with this new symmetry. The situation is similar to the late 60's 
when the hints in favor of the existence of a fundamental local $SU(2)$ 
symmetry of weak interaction were becoming compelling that eventually led 
to the triumphant gauge revolution in the 1970's. They were subsequently 
confirmed by the 
discovery of neutral currents, followed by the discovery of the W and Z 
boson. The two paths before us right now are (i) whether the B-L forces 
responsible for neutrino mass represent physics close to the gravity 
scale of $10^{18}$ GeV as exemplified, say by the grand unified theories 
or (ii) scales multi-TeV range far below the GUT scale with signals 
visible to the Large Hadron Colliders. One direct way to explore this new 
physics is to to carry out an experimental search for $n-\bar{n}$ 
oscillation.  As noted the discovery of $n-\bar{n}$ oscillation will 
have far reaching implications not only for the nature 
of forces and matter but will completely alter our thinking about such 
 cosmological issues as the origin of matter, nature of dark matter etc. 

The main points stressed in this review are that: (a)there are a wide 
class of models e.g. $SU(2)_L\times SU(2)_R\times SU(4)_c$ with type II 
seesaw, models based on TeV 
scale extra dimension models where $n-\bar{n}$ oscillation is in a range
accessible to currently planned reactor experiments and they will extend 
the existing limit on the strength of this process to an exciting range;
(b) Models that predict observable $n-\bar{n}$ oscillation provide a new 
way to understand the origin of matter where matter creation moment is 
``much'' later than currently contemplated models such as leptogenesis or 
GUT scale baryogenesis. More importantly, unlike the other mechanisms for 
baryogenesis, this mechanism can be tested in current collider 
experiments as well as neutrino experiments in some specific 
realizations.

  On the experimental front, searches for 
$n-\bar{n}$ transition using the proton decay type set-ups are back 
ground limited\cite{yuri} and cannot be used to push the limits for 
$\tau_{n-\bar{n}}$ too much beyond the present limit. Also an interesting 
class of theories that gives rise to $n-\bar{n}$ induced baryogenesis 
lead to new coloe sextet particles observable at LHC\cite{okada}.

Finally, it is also worth noting that there is a related phenomenon 
involving neutrons i.e. neutron-mirror neutron oscillation that can also 
be searched for in experimental set-ups similar to that searching for 
$n-\bar{n}$ oscillation\cite{bere}. This new class of oscillations have a 
much weaker experimental limit on its strength and could be improved by 
longer ``baseline'' searches being contemplated for $n-\bar{n}$ 
oscillation.

I would like to thank K. S. Babu for many discussions on theoretical 
aspects of the topic and  
 Y. Kamyshkov, G. Greene, Mike Snow and Albert Young for many discussions 
on the experimental prospects for $n-\bar{n}$ oscillation. I am grateful 
to K. S. Babu and Y. Kamyshkov for carefully reading the manuscript and 
suggesting improvements. This work is supported by the National Science
Foundation grant no. Phy-0652363.

\end{document}